# Novel detector systems for the Positron Emission Tomography

P. Moskal, P. Salabura, M. Silarski, J. Smyrski, J. Zdebik, M. Zieliński

Institute of Physics, Jagiellonian University, 30-059 Cracow, Poland

**Abstract:**
In this contribution we describe a novel solution for the construction of Positron Emission Tomograph. We present the device allowing for determination of the impact position as well as time and deph of interaction of the annihilation gamma quanta. The device is comprised of scintillation chamber consisting of organic scintillators surrounding the body of the patient. We discuss two possible solutions: (i) the tomograph built out of scintillator strips, and (ii) the tomograph built out of the scintillator plates. The application of the fast scintillators will enable to take advantage of the difference between time of the registration of the annihilation quanta. The invented method will permit to use a thick layers of detector material with the possibility of measuring the depth of the gamma quantum interaction (DOI) and the determination of their time of flight (TOF), and will allow for increasing the size of the diagnostic chamber without a significant increase of costs. The method is a subject of two patent applications [1,2] which are based on the techniques used in the particle physics experiments [3,4].

## Section 1: Introduction

Positron Emission Tomography (PET) is at present the most technologically advanced diagnostic method that allows for non-invasive imaging of physiological processes occurring in the body. It plays a fundamental and unique role both in medical diagnostics, as well as in monitoring of effects of therapy in particular in oncology, cardiology, neurology, psychiatry or gastrology. PET tomography constitutes also an effective tool to investigate the functioning of the brain. PET permits to determine the spatial and temporal distribution of concentrations of selected substances in the body. To this end, the patient is administered pharmaceuticals marked with radioactive isotope emitting positrons. Since the rate of assimilation of marked pharmaceuticals depends on the type of the tissues, sections of the diseased cells can be identified with high accuracy, even if they are not yet detectable via morphological changes. The method is proved to be extremely effective in particular in locating and diagnosing of cancer metastases.

All known matter including the body of the patient is built out of electrons, protons and neutrons. The PET uses the fact that the electron and positron annihilate while contact with each other and their mass is converted to energy in the form of gamma quanta. Most frequently these are two gamma quanta flying against each other along the line with an exactly defined energy equal to 511 keV. PET permits to locate the radioactive marker by the use of radiation detectors, allowing to reconstruct the direction of flight of annihilation quanta. Radiation detectors are usually arranged in layers forming a ring around the diagnosed patient (see Fig. 1). The set of reconstructed lines (referred to as Line of Response: LOR ) constitutes the basis for the reconstruction of the thomografic image which reflects the distribution of the density of the radiofarmacetic in the body of the patient. This technique allows to investigate the physiological processes involving the radiofarmaceutics since PET permits to obtain few tens of images within a minute.
The typical radiation dose in examination with PET amounts to about 7 mSv [5]. This is comparable with similar doses obtained by the patient in other diagnostics methods. It is in the order of the avarage yearly dosis due to the natural radiation sources as cosmic rays, concentrations of radon in the air etc. It has therefore no negative influence on the patient and even could have a positive implications in the functionning of the immune system due to the well established hormesis mechanism [6,7].

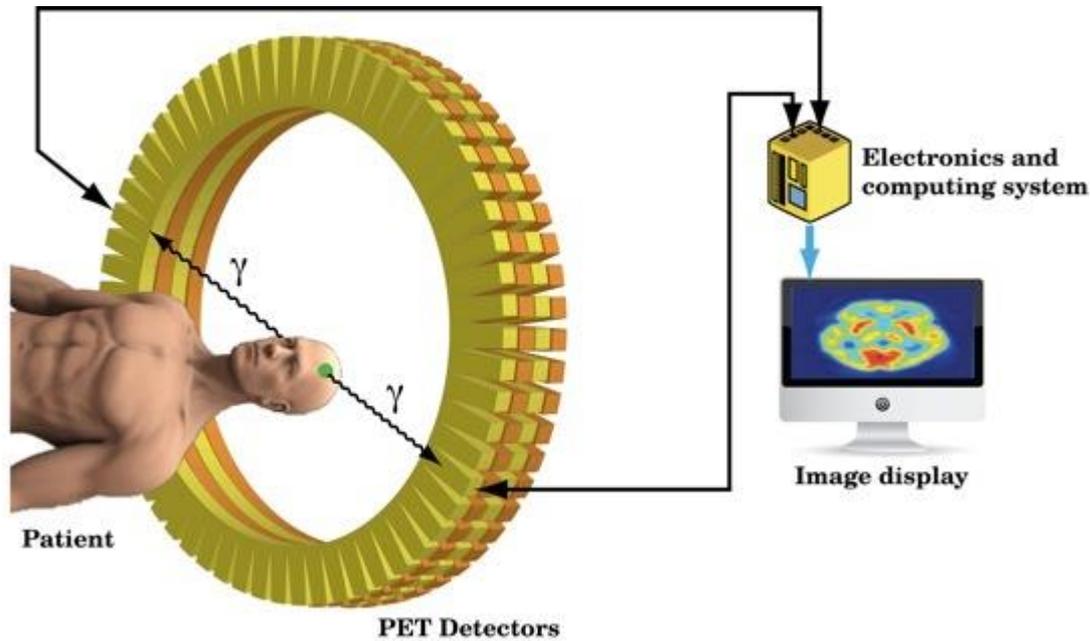

Fig. 1. Schematic illustration of the PET tomography.

A natural limitation of the sharpness of the PET image is given by the fact that positrons annihilates predominantly after its kinetic energy is decreased to the values close to zero which is typically few milimeters far from the nucleus from which it was emitted. As regards the detection technique among main factors limiting presently achievable accuracy are the problem of unknown depth at which gamma quantum reacts (referred to as DOI - depth of interaction), the problem of insufficient time resolution of non-organic detectors preventing them from the effective usage of the time difference between the arrival of the gamma quanta to the detectors (TOF - time of flight [8]) and, finally, present solutions are impractical to perform a tomographic image of the whole body at the same time.

In the following sections the presentation of the invented detector systems will be preceded by the description of the disadvantages in the tomografic image reconstruction caused by the unknown DOI and possible gains which could be achieved with the good resolution for TOF determination.

## Section 2: Depth of Interaction

Currently, all commercial PET devices use inorganic scintillator materials as radiation detectors (usually BGO cristals). The energy of gamma quantum hitting the scintilator can be transfered partially or entirely to an electron of the material, which then produces flash of lights through ionization and deexcitation of atoms or molecules of the scintillator. These flashes are then converted to electrical pulses by photomultipliers connected to the scintillators.

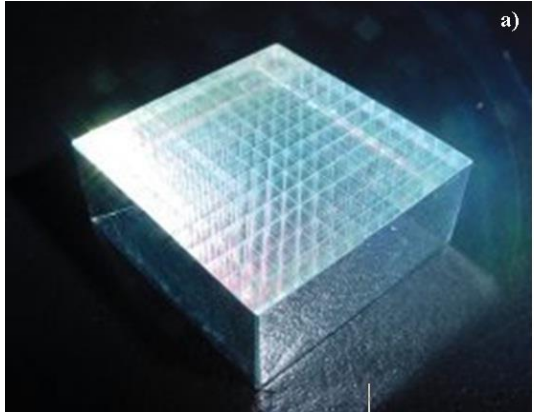 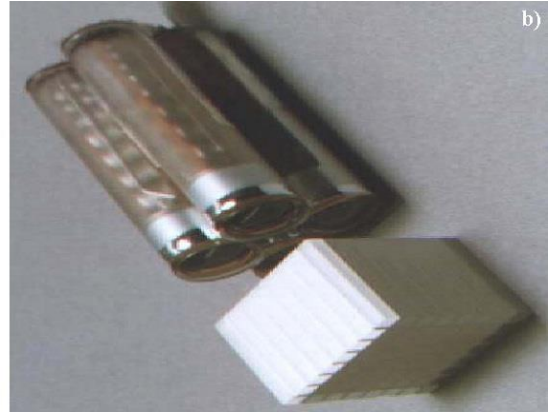

Fig. 2. a) Shape of a single scintillating module used to detect gamma quanta in PET scanners. b) Typical configuration of photomultipliers which register light from a single scintillating module used in PET scanners. The figures are adapted from [9,10].

As it was shown in Fig. 2a scintillating crystals, made usually in size of about 5cm x 5cm and with thickness of 2.5 cm, are additionally blazed into smaller pieces with dimensions of 0.5 cm x 0.5 cm separated form each other with reflecting material. The end of each scintillating module is connected to photomultipliers which convert light into electrical impulses (Fig. 2b). Distribution of the amplitude of this impulses permits to determine, with the accuracy equal to the size of the small unit, the position where the gamma quantum reacted. Therefore, in the further analysis, to determine the LOR line, one assumes that the quantum was absorbed in the middle of the unit. This assumption is one of the most important contribution limitting the resolution of PET images. Problem of unknown depth where the gamma quantum reacted is known in the literature as DOI (depth of interaction).

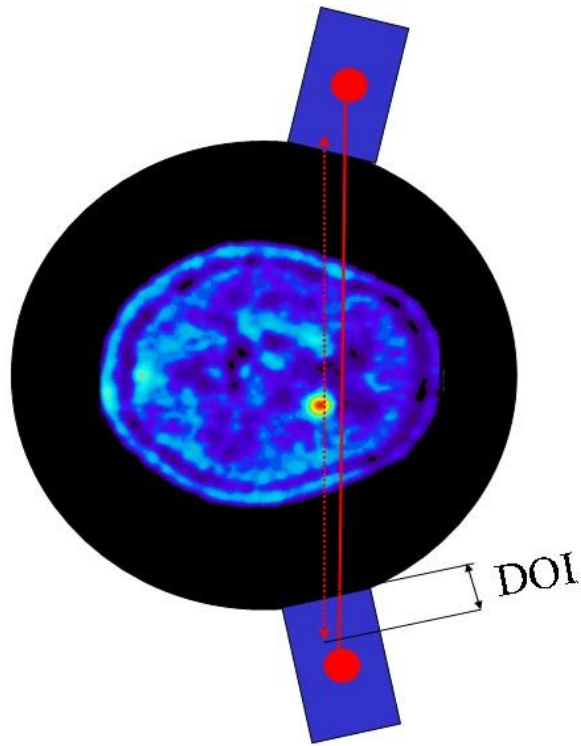

Fig. 3. Schematic illustration of the error in reconstruction of the LOR line made due to the unknown depth of interaction of the gamma quantum.

In Fig. 3 we present, in big simplification and exaggeration, the error in the determination of the LOR line made due to the unknown DOI. The dashed line shows the real path of flight of the gamma quantum, while the solid line represents the LOR reconstructed assuming that the signals emerged in the middle of the detection module. Distortions are the greater, the farther from the axis of the tomograph the annihilation occured, and the larger is the scintillator module. Therefore, the determination of the DOI could improve a lot the resolution far frrom the axis leading to better imaging of whole body. Moreover, it could allow to use a thicker scintillators improving the efficiency of the measurement. However, to our knowledge, at present none of commercially produced tomographs is able to measure the DOI.

## Section 3: Time-of-Flight

The other way to improve the resolution of the tomographic image is determination of the annihilation point on the LOR line based on measurements of the time difference between the arrival of the gamma quanta to the detectors. In the literature this technique is known as TOF (time of flight), and tomogrphs which use the time measurements are termed PET-TOF. In the idealized case, as it is shown in Fig. 4, measurement of the time difference between arrival of the quanta (t2-t1) could allow calculation of the annihilation point relative for example to the center of the LOR line denoted in Fig 4. as Δx. In practice, due to the finite resolution of the time measurement, it is possible to determine only a range along the LOR in which the annihilation occured which also improves the resolution of PET images.

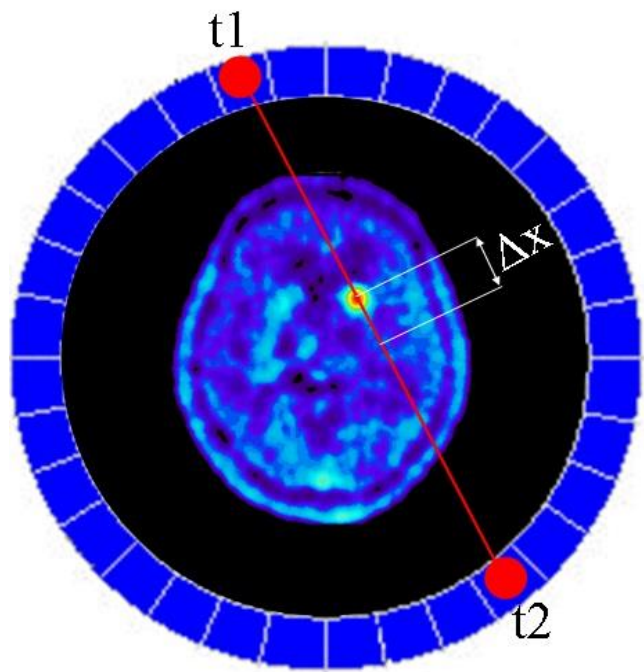

Fig. 4. The idea of PET-TOF; $\Delta x = (t2 - t1) \, c/2$.

Scientist have tried to use the time of flight of gamma quanta in the PET tomography since 1980 [11]. But so far nobody has obtained a significant improvement using TOF method, mainly because of the inorganic scintillators applied in PET scanners which give slow impulses. In 2008 a prototype made by SIEMENS achieved the time resolution of about 550 picoseconds which corresponds do the spatial resolution along the LOR line amounting to 8 cm [11]. About one order of magnitude better time resolutions can be achieved with organic plastic scintillators, which were so far not used due to low density and a small atomic numbers of the elements constituting the material. Fast plastic scintillators are composed mainly of carbon and hydrogen.

Small atomic number corresponds to small probability that gamma quanta transfer all its energy to the electron in the scintillator through the photoelectric effect. Moreover, small density implies a small efficiency for the detection of gamma quanta. The efficiency could be improved by increasing the thickness of the scintillator, but on the other hand it would decrease the resolution of the image due to DOI problem. However, novel methods presented in the next sections allow to increase the thickness of the detector and at the same time to determine the depth of the interaction of the registered gamma quantum. In addition due to the large solid angle covered by the new PET construction the decrease of the detector efficiency will be compensated by the increase of the accptance. A small efficiency for the photoelectric effect in organic scintillators worsen the image quality due to a low ability to distinguish between quanta reaching the detector directly and quanta rescattered in the body of a patient. This drowback will be compensated by (i) the selection of only these events for which the energy deposited in the scintillator corresponds to the range close to the maximum of energy which can be transfered to the electron via the Compton scattering process, and by (ii) taking advantage of the good timing of the organic scintillators allowing for the effective usage of the TOF technique.

In both below discussed methods the reconstruction of the interaction point of the gamma quantum in the scintillator material is reconstructed based predominantly on the time distribution of signals measured at various parts of the detector.

## Section 4. Novel Solution: Strip PET

In the "strip" PET the test chamber is formed from organic scintillator strips contracting the cylinder. Light signals from each strip are converted to electrical signals by two photomultipliers placed at opposite edges of each scintillator strip. The scheme of the single detector module is shown in Fig. 5. To determine the impact position of the gamma quantum the time difference between signals from both ends of the strip is used and the time of the interaction of the gamma quantum in the strip is determined as an arithmetic mean of the times measured on both edges of the scintillator.

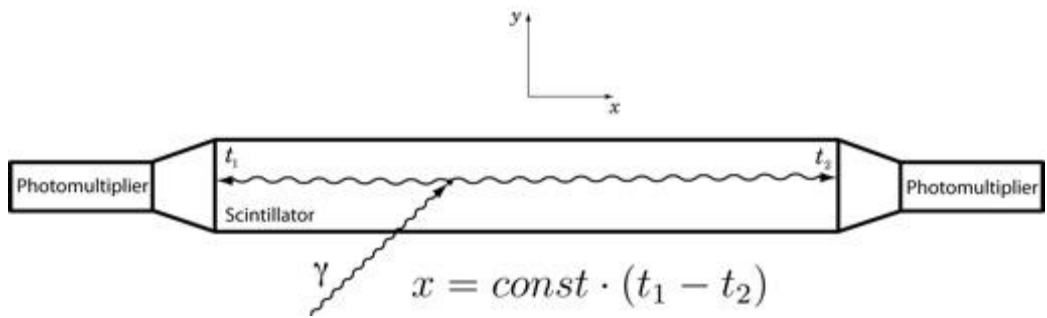

Fig. 5. Schematic view of a single detector module used in the "strip" PET. Position where the gamma quantum interacted can be determined from the difference between times measured at both edges of the strip.

The point of impact of a gamma quantum in a plane perpendicular to the axis of the strips can be determined from the position of a module which registered the signal, while the position along the scintillation chamber is determined using the difference between times measured in the front and rear photomultiplier. Energy of the electron colliding with gamma quantum is measured based on the amplitude of signals in the photomultipliers on both sides. Coincident registration of two gamma quanta allows to determine the line of response based on coordinates of reaction points reconstructed in both strips. The time of the reaction in each strip allows also to determine the annihilation point along the LOI based on the TOF method. The set of reconstructed LOR lines together with points of annihilation provides a tomographic picture.

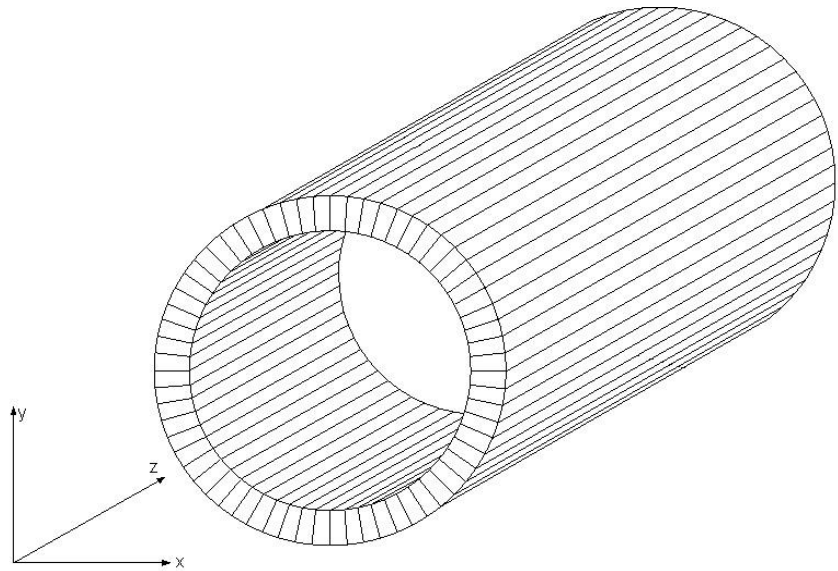

Fig. 6. Diagnostic chamber. Cylinder build out of scintillator strips.

## Section 5. Novel solution: Matrix PET

The Matrix-PET scanner would consist of organic scintillator plates, instead of currently used blocks of inorganic crystals. The plates could be set in many ways so as to cover the whole body of the patient, for example as it is shown in Fig. 7. The measurement of time and amplitude of light signals is carried out by photomultipliers matrix arranged around the chamber.

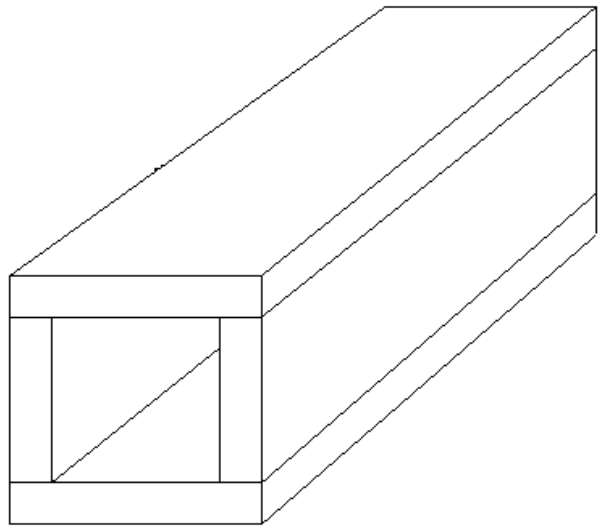

Fig. 7. Schematic view of the Matrix PET. Diagnostic chamber build out of scintillator plates.

The interaction point within the plane of the plate can be reconstructed based on both: (i) the distribution of the time of the signals from the side photomultipliers and (ii) distribution of amplitudes of the recorded signals. Such solution allows also to determine the depth at which the gamma quantum has been absorbed (DOI) on the basis of the distribution amplitudes of the signals from photomultipliers arranged on the sides (see Fig. 8 right). This feature allows to use thick plates without worsening of spatial resolution due to "the DOI problem" occurring in the current PET tomographs. Enlargement of the thickness enables efficient detection of gamma quanta using organic plastic scintillators, which are characterized by excellent time resolution, which is order of magnitude better in comparison with the fastest inorganic scintillators. Simultaneous registration of signals in blocks mounted in front of each other could enable determination of the gamma quantum path of flight. The time resolution through the use of plastic scintillators and as a result of simultaneous registration of the light signal by many photomultipliers would be much better than that achieved in the past PET scanners. Therefore, this solution would enable the usage of the TOF method permitting the determination of the annihilation point along the line of flight of annihilation quanta based on the time difference in reaching the different scintillation plates by the gamma quanta. Achieving TOF resolution of 50 picoseconds, possible when using organic scintillators and simultaneous measurement by many photomultipliers, would significantly simplify and hence make faster the reconstruction of the image.

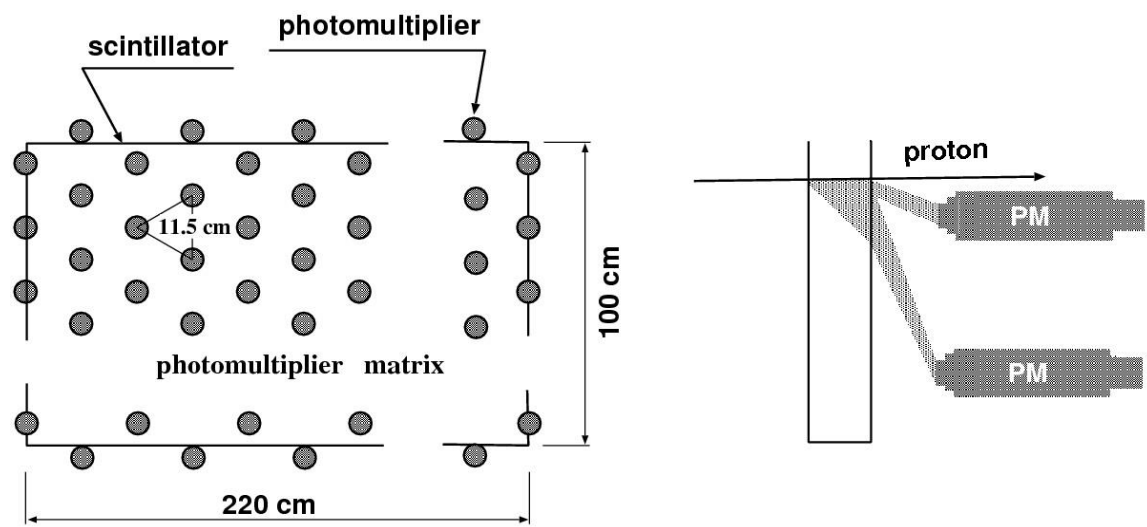

Fig. 8. (left and middle panels) Schematic view of the scintillation wall detector as used in the particle physics experiment COSY-11 [12]. The left panel demonstrates photomultiplier matrix together with the dimensions of scintillator wall, and in the middle pannel the principle of the light collection is shown. Right panel shows an edge of the scintillation chamber proposed for the Positron Emission Tomograph [2].

## Section 6. Summary

Both solutions Strip-PET and Matrix-PET offer an opportunity to obtain high time resolution and high acceptance, which are unaffordable in the current devices and which may translate into the precision of the annihilation point determination.

Proposed technology of construction of the PET tomograph allows the use of organic scintillators (plastics), which are relatively easy to produce in different shapes and large sizes, in contrast to currently used inorganic crystals. An ability of relatively easy extension of the size of the diagnostic chamber, is applicable particularly in "strip" PET where such extension does not entail an increase in the number of photomultipliers. This feature would decrease the construction costs of the PET scanners, which would also enable simultaneous imaging of the physiological processes throughout the whole body of the patient. Signals in currently used inorganic scintillators are much slower than the signals from the organic scintillators. Therefore, using fast organic scintillators (plastics) will also reduce the coincidence window giving the possibility to reduce accidental coincidences.

## Acknowledgements

The authors are thankful to Dr. Gabriela Konopka Cupiał and the team of the Centre for Innovation, Technology Transfer and University Development (CITTRU) for support in the investigations and for looking for companies and institutions interested in technology development, its testing and application. The work is also partly supported by The European Social Fund and Poland's Ministry of Treasury through the Scholarship for PhD Students „DOCTUS" and the „Małopolska Scholarship for PhD Students".